\documentclass[sn-nature]{sn-jnl}
\usepackage[utf8]{inputenc}
\usepackage[T1]{fontenc}
\usepackage{bbm,dsfont}
\usepackage{tikz}
\usepackage{graphicx}%
\usepackage{multirow}%
\usepackage{amsmath,amssymb,amsfonts}%
\usepackage{amsthm}%
\usepackage{array}
\usepackage{mathrsfs}%
\usepackage[title]{appendix}%
\usepackage{xcolor}%
\usepackage{textcomp}%
\usepackage{manyfoot}%
\usepackage{booktabs}%
\usepackage{algorithm}%
\usepackage{algorithmicx}%
\usepackage{algpseudocode}%
\usepackage{listings}%
\usetikzlibrary{quantikz}
\usepackage{soul}

\renewcommand{\ket}[1]{|#1\rangle} 
\renewcommand{\bra}[1]{\langle#1|} 
\newcommand{\kb}[2]{|#1\rangle\!\langle#2|} 
\newcommand{\tr}[1]{\mathrm{tr}\left[#1\right]} 
\newcommand{\id}{\mathbbm{1}} 
\newcommand{\A}{\mathsf{A}}
\newcommand{\B}{\mathsf{B}}
\newcommand{\C}{\mathsf{C}}
\newcommand{\D}{\mathsf{D}}
\newcommand{\I}{\mathcal{I}}

\DeclareMathOperator{\supp}{supp}
\newcommand{\diag}{\mathrm{diag}}
\DeclareMathOperator{\rank}{rank}

\begin{document}

\title{General measurements with limited resources and their application to quantum unambiguous state discrimination}

\author*[1,2,3]{\fnm{Daniel} \sur{Reitzner}}\email{daniel.reitzner@vtt.fi}
\equalcont{These authors contributed equally to this work.}
\author[2]{\fnm{Jan} \sur{Bouda}}
\equalcont{These authors contributed equally to this work.}

\affil*[1]{\orgdiv{Quantum Algorithms and Software}, \orgname{VTT Technical Research Centre of Finland}, \street{P.O.~Box 1000}, \postcode{FI-02044 VTT}, \country{Finland}}
\affil[2]{\orgdiv{Faculty of Informatics}, \orgname{Masaryk University}, \street{Botanick\'a 68a}, \postcode{602 00} \city{Brno}, \country{Czech Republic}}
\affil[3]{\orgdiv{RCQI, Institute of Physics}, \orgname{Slovak Academy of Sciences}, \street{D\'ubravsk\'a cesta 9}, \postcode{845 11} \city{Bratislava}, \country{Slovakia}}

\abstract{
In this report, we present a framework for implementing an arbitrary $n$-outcome generalized quantum measurement (POVM) on an $m$-qubit register as a sequence of two-outcome measurements requiring only single ancillary qubit. Our procedure offers a particular construction for the two-outcome partial measurements which can be composed into a full implementation of the measurement on any gate architecture. This implementation in general requires classical feedback; we present specific cases when this is not the case. We apply this framework on the unambiguous state discrimination and analyze possible strategies. In the simplest case, it gives the same construction as is known, if we opt for performing conclusiveness measurement first. However, it also offers possibility of performing measurement for one of the state outcomes first, leaving conclusiveness measurement for later. This shows flexibility of presented framework and opens possibilities for further optimization. We present discussion also on biased qubit case as well as general case of unambiguous quantum state discrimination in higher dimension.
}


\maketitle

\section{Introduction}

Measurements are one of the crucial elements of quantum theory. Compared to the usual notions of observables as self-adjoint operators, or of von Neumann measurements, there exists a more general description of measurements, the so-called positive operator-valued measures (POVMs). POVMs are defined by a set of positive operators summing up to the identity operator~\cite{Peres2006}. Being more general, POVMs outperform projective measurements for many tasks in quantum information theory, including quantum tomography~\cite{Renes2003}, unambiguous discrimination of quantum states~\cite{Bergou2010}, state estimation~\cite{Derka1998}, quantum key distribution~\cite{Bennet2002,Renes04Crypt,Nielsen2010}, information acquisition from a quantum source~\cite{Jozsa2003}, Bell
inequalities~\cite{Gisin96,Vertesi2010,tavakoli2021} or device-independent quantum information protocols~\cite{Acin2016,Gomez2016}.

Current quantum computation era, dubbed as \textit{Noisy intermediate-scale quantum} (NISQ) by John Preskill \cite{Preskill18}, is characterized by devices which still provide limited resources. Firstly, their size is limited only to a small number of qubits. Secondly, the number of computation steps (circuit depth) and their precision is still very limited by the decoherence effects in current technology, making the efficient usable number of qubits much smaller.

Final restraint is the limitation imposed on implementable quantum measurements --- current devices are tuned to perform only projective measurements in the computational basis. While projective measurements are the ideal result, in reality, the measurements are rather noisy as well. It could be noted that on the lowest level of physical qubits one can possibly tune up the measurements to perform an arbitrary measurement; in this paper, we are not addressing this topic.

Our aim is to provide a technique to implement POVM measurements with stated limited resources using gate formalism with focus mostly on the circuit depth/width trade-off. In addition, we do not concentrate on the noisy measurements \cite{Andersson12b,MaZiOs20}, the implementation is analyzed in the idealized scenario of perfect implementation and shows fundamental possibility of such realization and provides basic building blocks to do so.

Suppose we want to perform an $n$-outcome (POVM) measurement in a $d$-dimensional Hilbert space, $\A=\{A_j\}_{j=0}^{n-1}$. By a naive interpretation of Naimark theorem \cite{Peres2006}, one needs an ancillary Hilbert space of dimension up to $dn$. This is a single-step measurement procedure, where the measurement on the whole $dn$-dimensional space at some point provides complete outcome information; this, however, requires $\log_2 dn$ qubits and the decomposition into basic gates will increase the depth of the circuit considerably. So far the best approach needs an additional Hilbert space of the same dimension as is the dimension of the original Hilbert space irrespective of the number of outcomes of the measurement at question \cite{OsGuWiAc17}.

On the other end lies the result of Ref.~\cite{OsMaPu19}, where the spatial resources (system dimension) are exchanged for the decreased success probability. In the paper, the whole measurement is performed on the original $d$-dimensional Hilbert space, but it is successful only with probability $1/d$. The lowered success rate is in general inevitable.

On one extreme, simple dilation may need more resources than available, on the other extreme, measurements limited just to the original Hilbert space decrease probability of success and one does not have direct access to the measurement $\A$. In this case one can only reconstruct statistics by post-selecting obtained data, but per-shot relation of the outcomes to the measurement $\A$ cannot be interpreted directly. We would like to explore possibilities how to retain reasonable memory requirements (number of additional qubits), while not decreasing success probability. We also want to explore only technical possibilities and the paper will not discuss philosophy of what constitutes direct measurement of a POVM. 

Inspired by \cite{AnOi08}, we will concentrate on the next simplest model to the no-ancilla approach. We will use only a single ancillary qubit to provide us with a possibility of performing simple (two-outcome) measurements. We shall explore this option from a point of view determining practical ways of performing complex measurements as a sequence of simple ones as depicted in Fig.~\ref{fig:partitioning}. A similar approach using single ancilla was presented in \cite{SiMaOs2021} with the difference that our approach does not require post-selection. The price to pay in our case is a setup requiring deeper circuits and longer coherence times.

As noted, our approach is not entirely new and partial results can be found in some other works, most importantly \cite{AnOi08}, where the foundation for the sequential POVM implementation has been laid down. Other results contain some steps for constructions but they often lack generality. Quite often they are tailored for specific qubit implementations --- we present comparison on different approaches in a designated Sec.~\ref{sec:comparison}. In this paper we provide a general gate implementation procedure for any $d$-dimensional qubit-based system including the description of all steps of the process and all building blocks in one place. One can also view our results as partially following from more general approaches of single-qubit ancilla driven computation \cite{AnOiKaBrAn10,ShenEtal17}.

\subsection{Generalized measurements}

Generalized measurements, or \emph{Positive operator-valued measures (POVMs),} are a general way of describing measurements in quantum theory. In the finite-outcome case we are about to study, an $n$-outcome POVM $\A$ is represented by a set of operators $\A=\{A_j\}_{j \in [n]}$, where $[n]=\{0,1,\ldots,n-1\}$.
Operator $A_j$ corresponds to the outcome $j$; having state $\rho$ on which we perform measurement $\A$, outcome $j$ is obtained with probability $p(A_j|\rho)$ that is given by \emph{Born formula,} $p(A_j|\rho)=\tr{A_j\rho}$. This demands that the operator $A_j$ is positive semi-definite, $A_j\geq 0$; these operators are called \emph{effects}.

We also require that the probabilities sum up to one,
\[
1=\sum_{j\in[n]} p(A_j|\rho) = \sum_{j\in[n]}\tr{A_j\rho} = \tr{\rho\sum_{j\in[n]} A_j}.
\]
As this has to hold for all states $\rho$, it follows that the sum of the POVM effects equals identity,
\[
\sum_{j\in[n]} A_j = \id.
\]
Note, that von Neumann (projective) measurements are a special case of POVMs, as any projective measurement is described by a set of particular projections, which are also effects.

\begin{figure}
    \centering
    \includegraphics[scale=0.8]{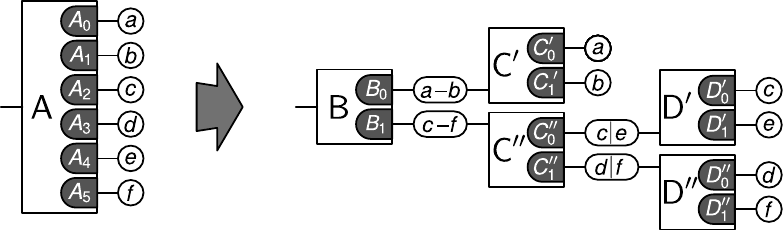}
    \caption{An example of a coarse-graining illustrating possible implementation of a POVM in a sequential way utilizing two-outcome measurements. Imagine measurement $\A=\{A_j\}_{j\in [6]}$ with six outcomes labeled $a$--$f$. Measurement $\B=\{B_0,B_1\}$ is a coarse-graining of $\A$ having two outcomes, one being a collective outcome for the labels $a$ and $b$ of the measurement $\A$ and the other one being a collective outcome for the labels $c$--$f$ of the measurement $\A$. The idea of the paper is to use such two-outcome coarse-grainings in a sequential way to perform the measurement $\A$. In this case, if the measurement $\B$ gives the label $a$--$b$, it is followed by the measurement $\C'$ giving a definitive answer $a$ or $b$. If, however, the $\B$-measurement gives the label $c$--$f$, it is followed first by the measurement $\C''$ and based on its outcome either $\D'$ or $\D''$ is performed and provides one of the definitive outcomes of the measurement $\A$.}
    \label{fig:partitioning}
\end{figure}

For the purposes of this paper, we also define the notion of \emph{coarse-graining} as exemplified in Fig.~\ref{fig:partitioning}.
Let us have a partition $P=\{P_k\}_{k\in[w]}$ (for some number of partitions $w$) of the outcomes of the measurement $\A=\{A_j\}_{j\in [n]}$, namely $P_k\subseteq [n]$ such that $\cup_{k} P_k=[n]$ and $P_j\cap P_k=\emptyset$ for all $j\neq k$. A coarse-graining is such a measurement $\B=\{B_k\}_{k}$ that composes outcomes according to given partitioning $P$, $B_k=\sum_{j\in P_k} A_j$. Later in this work we will restrict ourselves only to dichotomic coarse-grainings where there are only two partitions of the measurement.

We will also use the term \emph{fine-graining}, which is an opposite to coarse-graining, i.e., it corresponds to splitting of effects $B_k$ to sub-effects, providing a finer measurement. For example, having a measurement $\A=\{A_0, A_1, A_2, A_3\}$ a coarse graining can be a measurement $\B=\{B_0,B_1\}$, where $B_0=A_0+A_1+A_3$ and $B_1=A_2$. Measurement $\A$ is then a fine-graining of $\B$.

We also note that occasionally we will discern between an outcome and its label --- while the effects of a POVM are in this paper usually indexed by the elements of the set $[n]$, in general they can have assigned different labels (as can be seen in Fig.~\ref{fig:partitioning} or Fig.~\ref{fig:sequences}). This is in particular useful in cases when we need a more descriptive information about the outcome, as, e.g., in the case of collective outcomes. It can be useful also in the case of binary (dichotomic, or two-outcome) measurements that we use, when the particular effects are indexed by the measurement outcome value from the computational basis (either 0 or 1), but the labels provide interpretation of given outcome. This distinction will not always be followed in this paper and sometimes the labels will be the same as the outcomes. It should, however, be clear from the context what form is meant by respective labeling or indexing.

\subsection{Naimark dilation theorem}

In this subsection, we present a mathematical model of performing a POVM measurement by extending the given Hilbert space to a higher dimensional Hilbert space and subsequently performing von Neumann measurements on the larger space to put our result better into perspective.

Let $\{F_i\}_{i\in[n]}$ be a POVM acting on Hilbert space $\mathcal H_A$ of dimension $d_A$. Then there exists a projective measurement $\{P_i\}_{i\in[n]}$ acting on the Hilbert space $\mathcal H_{A'}$ of dimension $d_{A'}$ and an isometry $S:\mathcal H_A\longrightarrow \mathcal H_{A'}$ such that for all $i$
\begin{equation}
    F_i=S^{\dagger}P_i S.
\end{equation}
A naive (and inefficient) way to construct such projective measurement and isometry is to let $\mathcal H_{A'}=\mathcal H_A\otimes\mathcal H_B$, $P_i=I_A\otimes\ket{i}_B\bra{i}$, and
\begin{equation}
    S=\sum_{i\in[n]}\sqrt{F_i}_A\otimes\ket{i}_B.
\end{equation}
This construction, however, requires a system of large dimension, and $d_{A'}=nd_A$.
This approach to POVM can be turned into physical implementation by extending the isometry $S$ to a unitary operation $U$ that fulfills
\begin{equation}
    S=U(I_A\otimes\ket{0}_B).
\end{equation}

A more dimension--efficient approach was designed by Peres \cite{Peres2006}, where the construction requires dimension
\[
d_{A'} = \sum_{i\in[n]} \rank{F_i}.
\]
In \cite{OsGuWiAc17}, the authors provide another construction of dilation requiring an ancillary system of the same dimension as the original system irrespective of the number of outcomes.

In this work we will similarly extend the studied system but only by a qubit system. This dilation to a qubit, however, limits possibilities for our intended measurements. Namely, we cannot expect to be able to perform a measurement with more than two outcomes (on the ancillary qubit system). This, in turn, defines a way, how we will approach the problem of measuring more outcomes --- we will look at the possibility of splitting the measurements into a sequence of two-outcome measurements.

Note that particularities of used device may play role in the size of the required ancilla. While we discussed the necessity of one ancillary qubit, this is only in the case when it can be dynamically reset. If this is not the case, one will need single-qubit ancilla for each measurement in the sequence.

\subsection{Measurements with state changes}

POVMs describe measurements only from the perspective of outcomes and their probabilities. They do not, however, describe what happens to the measured state. As in the sequential implementation, we want to reuse the system for fine-graining, we need to be able to describe also measurements with a state change.

In the case of von Neumann measurements, the change to the state $\rho$ when outcome $j$ corresponding to projector $P_j$ is measured, is given as $\tilde\rho_j = P_j\rho P_j$. The operator $\tilde\rho_j$ is not normalized, and its normalization provides both the outcome state
\[
\rho_j = \frac{P_j\rho P_j}{\tr{P_j\rho}}
\]
and the probability of getting this outcome, $p(P_j|\rho)=\tr{P_j\rho}$.

For a general POVM, we will describe these measurement-induced state changes as \emph{instruments}. An instrument $\I$, corresponding to the measurement $\A$ is a set of completely positive trace non-increasing maps $\I=\{\I_j\}_j$ such that
\begin{equation}
\label{eq:instrument}
\tr{\I_j(\rho)} = \tr{A_j\rho}
\end{equation}
which has to hold for all states $\rho$. The positivity of $A_j$ translates to the requirement that $\I_j$ is completely positive, while the summation condition for $\A$ translates to the requirement that the sum of $\I_j$'s is a channel (completely positive trace preserving map) which implies the trace non-increasing property on the particular $\I_j$'s. As before, operators $\tilde\rho_j=\I_j(\rho)$, representing what happens to state $\rho$ when outcome $j$ is observed, are not normalized, with probability $p(j|\rho)$ of obtaining the outcome being the normalization factor, i.e., the outcome state is given as
\[
\rho_j = \frac{1}{p(A_j|\rho)}\I_j(\rho) = \frac{\I_j(\rho)}{\tr{\I_j(\rho)}} = \frac{\I_j(\rho)}{\tr{A_j\rho}}.
\]
We will use \emph{measurement} as the name for both POVMs and instruments; however, it should be clear which one is used.

An important thing to note is that
while for von Neumann measurements the presented state change is unique, in the general case of POVMs, the choice is not unique. 
Different choices can affect the state in different ways and, in particular, can lead to various degrees of state disturbances. For example, the instrument
\[
\I_j(\rho) = \tr{A_j\rho}\omega
\]
for some state $\omega$ destroys the original state $\rho$ completely.
It is therefore natural to try to find the least disturbing choices, especially, when the resulting state is to be used later.

In \cite[Prop.~5.17]{HeZi11}, it was shown that the so-called L\"uders measurements (or instruments), given by the prescription
\begin{equation}
\label{eq:luders}
\I_j(\rho)\equiv\mathcal L_j(\rho)=A_j^{1/2}\rho A_j^{1/2},    
\end{equation}
are the least disturbing in the following sense: any measurement can be realized as a L\"uders measurement followed by some outcome-dependent state change. This makes them a straightforward choice in our endeavor.

Note that there is a number of possible additional criteria that may give preference to other than L\"uders instruments, such as hardware specifics, noise considerations, or preferences following from a particular measurement goal. This is, however, beyond the scope of this paper.

\section{Measuring with limited resources}

As we noted before, current quantum devices provide us with highly limited resources. If a desired measurement is more complicated, these resources might not even allow us to implement it. A straightforward idea is to split the measurement into a sequence of binary measurements as depicted in Fig.~\ref{fig:partitioning}. While in the classical world such action bears no problems, in the quantum case we know, that every measurement disturbs measured state. A question arises, whether it is possible to devise a general procedure that would allow us to implement the measurement in such a sequential way. The question has two parts, (i) whether on general level such splitting is theoretically possible, and (ii) if so, whether this procedure is implementable.

The answer to the first question has been to large extent provided in \cite{AnOi08}. In this paper, we present a slightly different way of obtaining the result, and later, we apply this procedure to study unambiguous state discrimination. We start by showing that the L\"uders measurements allow for fine-graining of results, answering point (i). Then we show how to implement qubit-assisted measurements, what in turn allows us to implement this procedure on current quantum devices based on qubit registers. This shall answer point (ii).

\subsection{POVM as a sequence of binary measurements}

Let us consider a measurement $\A$ and its coarse-graining $\B$. We will consider only a two-outcome coarse-graining as (i) we want to study the possibilities of single ancillary qubit that distinguishes only two outcomes, and (ii) the analysis of the procedure to higher number of outcomes is straightforward. Let us have $\B=\{B,\id-B\}$ and $Q$ being the subset of outcome indices of measurement $\A$ that defines $B$, i.e., $B=\sum_{j\in Q} A_j$.

Let us now take the case when the effect $B$ was measured on the input state $\rho$ and the state change is described by L\"uders measurement from Eq.~(\ref{eq:luders}). The (unnormalized) state now is $\tilde\rho=B^{1/2}\rho B^{1/2}$. If we now want to fine-grain the results to obtain information about outcomes from $Q$, we cannot perform measurement $\A$ on the state $\tilde\rho$ any more --- we need to design a new measurement adjusted for the fact that the previous measurement $\B$ has already been done, particular outcome corresponding to the effect $B$ has been obtained, and the state has changed.

In fact, what we want is to find measurement $\A'=\{A'_j\}_{j\in Q}$ such that the following holds
\begin{equation}
    \label{eq:subseqcond}
    \tr{\tilde\rho A'_j}=\tr{\rho A_j}.
\end{equation}
Expanding the left hand side we see
\[
\tr{\tilde\rho A'_j}=\tr{B^{1/2}\rho B^{1/2}A'_j}=\tr{\rho B^{1/2}A'_jB^{1/2}}.
\]
Since this has to be equal to $\tr{\rho A_j}$ for all $\rho$, we obtain condition for $\A'_j$ stating that
\begin{equation}
    \label{eq:Aj_prime_conditions}
A_j=B^{1/2}A'_j B^{1/2}.   
\end{equation}
Let us take the Moore-Penrose pseudoinverse of $B^{1/2}$, which we denote simply as $B^{-1/2}$. We see that $A'_j$ given by
\begin{equation}
    \label{eq:updatedmeasurement}
    A'_j=B^{-1/2}A_j B^{-1/2}
\end{equation}
satisfies equation \eqref{eq:Aj_prime_conditions}. Since $B\geq 0$, we also have $A'_j\geq 0$. It remains to verify the normalization,
\[
\sum_{j\in Q} A'_j=\sum_{j\in Q}B^{-1/2}A_j B^{-1/2} = B^{-1/2}BB^{-1/2}=\id_B,
\]
where $\id_B$ is the identity (projection) on the support of $B$. We need not be concerned with the rest of the Hilbert space, as for the operator supports we have $\supp A_j\subseteq\supp B$ and also $\supp\tilde\rho\subseteq\supp B$. This means that while the transformed state loses information outside the $\supp B$, the subsequent measurements anyway act only within $\supp B$. So we see that Eq.~(\ref{eq:updatedmeasurement}) is sufficient to define the subsequent measurement that fulfills the condition from Eq.~(\ref{eq:subseqcond}). As for the particular implementation, $\id-\id_B$ can be defined as an additional outcome in order for $A'$ to be a POVM on the whole Hilbert space, but the probability of obtaining this outcome will be zero. Alternatively, one can supplement the effects of $\A'$ by parts from the orthocomplement of $B$ so that they sum up to identity. This will not change the probabilities, as $\tilde\rho$ does not have support in this part of the space.

We can consider a number of possible strategies how to partition a particular measurement into a measurement tree as exemplified in Fig.~\ref{fig:partitioning}. The role in their usefulness might have hardware specifics, noise considerations, or preferences following from a particular measurement goal. Disregarding these considerations, two partitioning strategies stand out.

\begin{figure}
    \centering
    a)\includegraphics[scale=0.8]{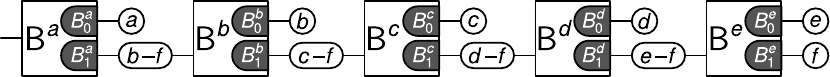}\\[8mm]
    b)\includegraphics[scale=0.8]{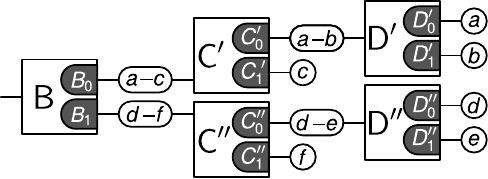}
    \caption{Examples of possible measurement procedures. Figure a) depicts an outcome-decreasing procedure representing an overall measurement $\A$ from Fig.~\ref{fig:partitioning}, where every measurement eliminates one outcome. In this case measurement eliminating label $x$ is denoted by $\B^x$. Figure b) shows a binary-search procedure of the same measurement $\A$ in which the number of possible outcomes is (roughly) halved in every step. Dark regions correspond to measured effects and circled outputs represent labels of given measurement.}
    \label{fig:sequences}
\end{figure}

\paragraph{Outcome-decreasing procedure:} In every step, we try to rule out one of the labels. Having a measurement $\A=\{A_j\}_{j\in [n]}$, in step $j$ we perform measurement $\B^j$  deciding between the outcome $j$ corresponding to $A_j$ and outcomes corresponding to the effects $A_{j+1},\dots,A_n$. If outcome $j$ is obtained, the measurement process can be terminated, since a definitive answer is obtained. This effectively means that conditioning on the previous outcomes is not required --- if definitive answer is obtained, the measurement procedure can continue as planned, but we can disregard the results obtained after the definitive answer. This implies that static circuits are enough for the implementation. The drawback of this procedure is large resulting circuit depth, as one needs to perform $n-1$ consecutive steps of the measurement process. The procedure is depicted in Fig.~\ref{fig:sequences}a.

\paragraph{Binary-search procedure:} In this procedure we split the outcomes of current measurement in (roughly) half and based on given outcome we choose the next measurement to be done. This procedure is depicted in Fig.~\ref{fig:sequences}b and was presented already in \cite{AnOi08,RoDeMaGi17}. Compared to the outcome-decreasing procedure, it is more efficient in circuit depth, as the number of steps one needs to make is roughly $\log_2 n$. The price to pay is that one needs to be able to condition measurements to be done on the previous outcomes. This option is becoming available in current quantum devices, but it still might have unreasonable time demands or low quality.\\

There are, naturally, many other options how to approach the coarse-graining procedure.
There seem to be only few special cases when the non-adaptive approach is possible (outcome-decreasing procedure or the implementation of symmetric informationally-complete POVMs as, e.g., in \cite{GaReSa23}), as in general the procedures need to be adaptive (requiring dynamic circuits), where the later measurements depend on previous results.

\subsection{Qubit implementation of L\"uders measurements}

Remaining question now is, whether there is a simple and an efficient realization to an arbitrary binary measurement. In this part, we shall show that using one ancillary qubit, we can represent any two-outcome measurement $\B=\{B,\id-B\}$ as a rotation of the original system to the eigenbasis of $B$, followed by a controlled unitary with the ancillary qubit as the target, and finalized by a measurement on the ancillary qubit and rotation of the system back. We will further assume that the successful measurement outcome in the $\B$ measurement is 1 while the outcome 0 corresponds to the effect $\id-B$.

\begin{figure}
    \centering
    a)\input{img/generalconst}\\
    b)\input{img/CNOTconst}\\[2mm]
    \caption{Coupling scheme for binary measurements. a) In general setting the (not necessarily single-qubit) state $\rho$ is coupled by a unitary $U$ to an ancillary qubit system prepared in the state $\ket{0}$, which is measured in the computational $z$ basis afterward. Given outcome of the measurement is $j$, the (unnormalized) output state is $\tilde\rho_j$. b) The scheme in the case of L\"uders measurements can be decomposed into a rotation of $\rho$ to the eigenbasis of $B$ by $U_B$ (and back at the end). The rest of the coupling unitary is a general control operation of the form $V=\sum_k \kb{k}{k}\otimes V_k$. When some $V_k$ is identity, it can be excluded from the circuit.}
    \label{fig:generalconstruction}
\end{figure}

Let us present a detailed description. We start with a general coupling construction as in Fig.~\ref{fig:generalconstruction}a, where the original state (Hilbert space $\mathcal H_1$) is coupled to an ancillary single-qubit state (Hilbert space $\mathcal H_2$), which is measured afterward. Without loss of generality, we assume that the ancillary qubit is prepared in the state $\ket{0}$ and its measurement is in the computational ($z$) basis. In fact, this is compatible with the default settings of most contemporary quantum computers. This construction gives us the following conditions for $U$ acting on $\mathcal H_1\otimes\mathcal H_2$; for the two outcome cases (producing either state $\tilde\rho_1$ or $\tilde\rho_0$) we have:
\begin{align*}
    B^{1/2}\rho B^{1/2} & = {}_2\bra{1}U\ket{0}_2 \rho\, {}_2\bra{0} U^\dagger\ket{1}_2,\\
    (\id-B)^{1/2}\rho (\id-B)^{1/2} & = {}_2\bra{0} U\ket{0}_2\rho\, {}_2\bra{0} U^\dagger\ket{0}_2,
\end{align*}
where we explicitly use indexing marking the original system (1) and the ancillary qubit (2) where necessary. These two equations are, in particular, implied\footnote{We do not study general conditions as we are satisfied with any solution.} by the following conditions:
\begin{equation}
\label{eq:U_conditions}
{}_2\bra{1} U\ket{0}_2 = B^{1/2}\ \ \text{and}\ \ 
{}_2\bra{0} U\ket{0}_2 = (\id-B)^{1/2}.
\end{equation}
For the moment we leave open the question whether for each $B$ it is possible to construct a unitary $U$ satisfying these conditions.

Since $B$ is an effect, it can be diagonalized (in the computational basis) by a unitary transformation we denote $U_B$. This unitary diagonalizes at the same time both $B^{1/2}$ and $(\id-B)^{1/2}$. By emphasizing the diagonal form by the corresponding lower index, we have
\begin{align*}
    B^{1/2}_\diag &= U_B B^{1/2} U_B^\dagger 
    = U_B\, {}_2\bra{1} U\ket{0}_2\, U_B^\dagger
    = {}_2\bra{1} (U_B \otimes \id) U (U_B^\dagger \otimes \id) \ket{0}_2, \\
    (\id-B)^{1/2}_\diag &= U_B (\id-B)^{1/2} U_B^\dagger 
    = U_B\, {}_2\bra{0} U\ket{0}_2\, U_B^\dagger
    = {}_2\bra{0} (U_B \otimes \id) U (U_B^\dagger \otimes \id) \ket{0}_2.
\end{align*}
Denoting by $V=(U_B \otimes \id) U (U_B^\dagger \otimes \id)$, which is unitary if (and only if) $U$ is unitary, we can write it in the tensor product $\mathcal H_2\otimes \mathcal H_1$ (note that we swapped the order of the Hilbert spaces in the tensor product to achieve a more comprehensible form)
\begin{equation}
\label{eq:halfV}
V = \begin{pmatrix}
(\id-B)^{1/2}_\diag & V_{01} \\
B^{1/2}_\diag & V_{11}
\end{pmatrix}.
\end{equation}
where $V_{01}$ and $V_{11}$ are unknown submatrices that we aim to complete in such a way that $V$ is unitary.
We will proceed to show that for each measurement $\B$ it is possible to construct such a unitary matrix $V$. This will also show that $U$ satisfying conditions \eqref{eq:U_conditions} exists.

Denoting columns of $V$ as $v_k$, it is easy to see that for $j\neq k$ in the known part we have $v_j^\ast v_k=0$ as both matrices $B^{1/2}_\diag$ and $(\id-B)^{1/2}_\diag$ are diagonal. Denoting the (real) eigenvalues of $B$ as $\lambda_k \in [0;1]$, for the column norm we have
\begin{align*}
v_k^\ast v_k &=
\left[ (1-\lambda_k)^{1/2} \right]^\ast (1-\lambda_k)^{1/2} +
\left( \lambda_k^{1/2} \right)^\ast \left( \lambda_k^{1/2} \right)
= (1-\lambda_k) + \lambda_k = 1.
\end{align*}
So we see that the left part of $V$ fulfills the conditions for unitarity. It remains to find $V_{01}$ and $V_{11}$ such that all columns of $V$ are orthonormal. There is freedom in the choice, but we can choose
\[
V_{01}=-B^{1/2}_\diag\qquad\text{and}\qquad V_{11}=(\id-B)^{1/2}_\diag.
\]

If we now write the matrix $V$ in the original tensor order $\mathcal H_1\otimes\mathcal H_2$, it has a block-diagonal structure which is easily interpreted as a controlled operation of the form
\begin{equation}
\label{eq:ctrlcoupling}
V = \sum_k \kb{k}{k}\otimes V_k
\end{equation}
with $2\times 2$ matrices
\begin{equation}
\label{eq:Vk}
V_k = \begin{pmatrix}
(1-\lambda_k)^{1/2} & -\lambda_k^{1/2}\\
\lambda_k^{1/2} & (1-\lambda_k)^{1/2}
\end{pmatrix},
\end{equation}
where $k\in[d]$.
Note that if $\rank B = r$, then $r$ of these matrices are non-trivial, while the rest of them equals $\id_2$; implementation then requires $r$ controlled operations for the $r$ non-trivial matrices.

To sum up, the procedure from Fig.~\ref{fig:generalconstruction}a can in particular be constructed as in Fig.~\ref{fig:generalconstruction}b, in which we first rotate the original state to the $B$-eigenbasis, then perform controlled operations with an ancillary qubit as target, and finally measure the ancillary qubit and rotate the original system back from the $B$-eigenbasis. Presented construction is general and works in any dimension $d=2^m$ of the original system.

In the next part dealing with application to unambiguous state discrimination, we will consider the simplest case, when the original system is a qubit.
In such case, we can rewrite this controlled operation as
\begin{equation}
V = \kb{0}{0}\otimes V_0 + \kb{1}{1}\otimes V_1
= (\kb{0}{0}\otimes\id + \kb{1}{1}\otimes V_1V_0^\dagger)
(\id\otimes V_0),
\label{eq:coupling}
\end{equation}
which can be realized as a composition of a unitary transformation $V_0$ on the ancillary qubit and a standard qubit controlled-$(V_1V_0^\dagger)$ operation (see Fig.~\ref{fig:qubitconstruction}).

\begin{figure}
    \centering
    \input{img/qubitconst}
    \caption{The measurement scheme can be simplified even further in the case of a qubit system when the general controlled operation can be constructed as a composition of $V_0$ on the ancilla followed by a standard controlled-$(V_1V_0^\dagger)$ operation.}
    \label{fig:qubitconstruction}
\end{figure}

\subsection{Final overview of the procedure}

Previous steps provide a detailed description of the particularities of the implementation procedure. Now we provide a complete and cohesive view on the procedure to give a higher-level overview of how it works. The work flow is as follows:
\begin{enumerate}
    \item Choose measurement $\A$ for implementation.
    \item Decide on binary-tree coarse-graining of $\A$.
    \item Compute corresponding measurements for each partial step.
    \item For each partial dichotomic measurement $\B=\{B,\id-B\}$ compute diagonalization unitary $U_B$, and $U_B^\dagger$ and the set of coupling unitaries $V_k$ using Eq.~(\ref{eq:Vk}).
    \item Build the circuit stitching constructions from Fig.~\ref{fig:generalconstruction}b for particular measurements.
\end{enumerate}

The first two steps are to be chosen based on demands of the application and can reflect also specific considerations based on the QPU design and its calibration. How to choose coarse-graining is not the aim of this paper. Steps 3 a 4 are tasks from linear algebra relying on the ability to perform spectral decomposition of used operators. In the simplest tasks one can use analytic approach, but in general the decomposition complexity is of the order $O(d^3)$ where $d=2^m$ is the dimension of the $m$-qubit Hilbert space. For small problems of the NISQ era this is easily computable.

Practically demanding is the last step which in general requires implementation with the possibility of evaluation of circuits conditioned on previous outcomes. These possibilities are not generally available on current devices, and when implemented, they are not yet suited for current implementation. Therefore, at present it is necessary to ease of the demands that this paper sets. We see a few possibilities how to do that: (i) find use cases where the conditioning is not needed, such as quantum state discrimination, or (ii) use outcome-eliminating coarse-graining that can be evaluated without conditioning, or (iii) create a set of circuits implementing possible paths. The last point cannot be recommended as it removes the faithfulness of the measurement (realization requires post-selection), while maintaining  large depths of the circuits. However, we believe that the progress in the development of QPUs will lead to improvement of intermediate measurements.

\subsection{Comparison to other works}
\label{sec:comparison}

Let us now discuss the relation of our approach to other works. As mentioned in the introduction, the approach is not entirely new. However, we present the implementation in its entirety, from the theoretical analysis of the procedure, down to providing elementary building blocks for the implementation. To better describe how current paper fits into known results, we provide a list of relevant works in Tab.~\ref{tab:literature}.

The top part of the table lists works in the direction of sequential POVM implementation. Our approach is inspired by \cite{AnOi08} where a solid theoretical background for the coarse-grained procedure was given, but without particularities for the implementation. In addition, the presentation was using binary coarse-graining. Similar approach was used also in \cite{RoDeMaGi17}, where the authors focused on quantum state discrimination. Their results are not universal, but rather hard-tailored to the task. In \cite{YoBa19} the authors presented a constructive way for implementing POVMs, but only on a single qubit. Their realization requires $O(\log_2 n)$ ancillary qubits.

Some of the ideas for sequential implementation of POVMs is present also in papers practically utilizing the measurements but as we can see in Tab.~\ref{tab:literature} (bottom part), most of these are limited to one or two qubits. An exception is \cite{Fischer17} that works in any dimension, but is specifically designed for performing so-called informationally complete POVMs. All these implementations are platform specific and only hint at general implementation at best.

There are also other POVM implementation techniques that are based on different paradigms. All these implementations are suitable for POVMs of any dimension and any number of outcomes. In particular, in \cite{OsGuWiAc17} one of the results states that a POVM in a $d$-dimensional Hilbert space can be performed as a specific Naimark-type dilation with the ancillary Hilbert space of the same dimension as the original system. This work also provides a constructive proof that can be adapted for practical implementation. In references \cite{OsMaPu19, SiMaOs2021} a different approach is used with \cite{SiMaOs2021} requiring additional qubit. Both approaches provide shallow circuits, but the price to pay is the probabilistic nature of the implementation that requires post-selection. This means that these implementations do not represent desired POVMs faithfully. Finally, in reference \cite{PiZaBaMa23} the authors present a construction requiring $O(\log_2 n)$ ancillary qubits where the measurement is mapped to a problem of a particular state preparation which is measured on ancillas that provide the measurement outcome.

\begin{table}
    \centering
    \begin{tabular}{p{9.5em}p{2em}p{1.6em}p{1.6em}p{9em}p{11em}}
        \textbf{Paper} & Const. & $m$ & $n$ & \textbf{Additional resources} & \textbf{Description} \\
        \hline
        Andersson (2008) \cite{AnOi08} & No & any & any & 1 qubit with reset & sequential (binary)\\
        Rosati (2017) \cite{RoDeMaGi17} & No & any & any & 1 qubit with reset & sequential (binary), specific for USD\\
        Yordanov (2019) \cite{YoBa19} & Yes & 1 & any & $O(\log_2 n)$ & sequential (outcome-decreasing) \\
        \textit{this paper} & Yes & any & any & 1 qubit with reset & sequential (general)\\
        \hline
        Oszmaniec (2017) \cite{OsGuWiAc17} & Semi & any & any & Hilbert space of the dimension of $d$ & Naimark dilation \\
        Oszmaniec (2019) \cite{OsMaPu19} & Yes & any & any & time (post-selection) & success prob.~$p=1/d$ \\
        Singal (2021) \cite{SiMaOs2021} & No & any & any & time (post-selection), 1 qubit & constant success prob.\\
        Pinto (2023) \cite{PiZaBaMa23} & Yes & any & any & $O(\log_2 n)$ & state-preparation\\
        \hline
        Ahnert (2005) \cite{AhPa05} & Yes & 1 & any & --- & photonic, similar to outcome reduction\\
        Ahnert (2006) \cite{AhPa06} & Yes & 2 & $\sim 4$ & --- & photonic\\
        Ota (2012) \cite{OtAsNo12} & Yes & 1 & any & --- & linear optical and solid-state qubits\\
        Dressel (2014) \cite{DrBrKo14} & Yes & 1 & 2$+$ & 1 qubit with reset & superconducting, explicit for $n=2$, for larger $n$ hinted at sequential with outcome reduction\\
        Bian (2015) \cite{Bian15} & Yes & 1 & any & --- & rank-one POVMs, quantum-walks optical\\
        Zhao (2015) \cite{Zhao15} & Yes & 1 & any & --- & quantum-walks optical\\
        Fischer (2017) \cite{Fischer17} & Yes & any & $4^m$ & embedding in qudit space & superconducting, informationally-complete POVMs\\
        Singh (2022) \cite{SiArSa22} & Yes & any & any & --- & linear optical, sequential with outcome reduction\\
        \hline
    \end{tabular}
    \caption{Overview of related works. For the columns description, "Const." stands for whether the paper has a constructive design, $m$ denotes the number of qubits of the system (dimension of the Hilbert space is $d=2^m$) and $n$ denotes the number of outcomes of implemented POVM. Upper part of the table lists articles focusing on the theory of sequential POVM implementations, middle part lists articles with focus on implementations using different paradigms and the bottom part of the table lists articles with focus on particular platform implementation.}
    \label{tab:literature}
\end{table}

We conclude that while the literature is rather rich with implementations of POVMs, they are mostly limited in one or another way. Some results are just proof-of-the-principle, some are limited in the scope. In this paper, we present a complete and faithful construction for POVMs on any number $m$ of qubits (i.e., the dimension is $d=2^m$) and having an arbitrarily (finitely) many outcomes $n$.

\section{Quantum unambiguous state discrimination as a sequential measurement process}

\subsection{Quantum unambiguous state discrimination}

\emph{Quantum state discrimination} is a task in which we are provided a state from a set of states $\{\rho_j\}_{j\in[w]}$ with probabilities $\{p_j\}_{j\in[w]}$ and our task is to determine, which state was presented to us. Due to the particularities of quantum mechanics, this task is not as straightforward as in the classical case --- in quantum theory one cannot distinguish non-orthogonal states perfectly. This task is therefore of high importance.

A particular situation of \emph{unambiguous state discrimination} was introduced in \cite{Dieks1988, Ivanovic1987}. In this setting we want to distinguish particular states without errors, i.e., if we are given a definite answer about the state, it needs to be correct. The price to pay for this requirement is the necessity for an \emph{inconclusive} outcome. When we obtain this result, we cannot say anything particular about the presented state.

This particular task has many extensions, but for the sake of exemplifying the framework from the previous section we will deal with the most basic setting of being presented with two pure qubit states $\{\ket{\psi_1},\ket{\psi_2}\}$ with equal probabilities. Our task is to determine, which state was given to us.

Let us first denote $\ket{\psi_j^\perp}$ as states perpendicular to $\ket{\psi_j}$ and $P_j$, $P_j^\perp$ as the corresponding projectors.
The unambiguous state discrimination measurement $\A$ has three outcomes 1, 2, and ? (for the inconclusive outcome) with corresponding effects,
\[
A_1 = \lambda P_2^\perp,\quad
A_2 = \lambda P_1^\perp,\quad
A_? = \id - A_1 - A_2.
\]
The choice for effects $A_1$ and $A_2$ is logical, as it tells us that the presented state is not the other one. Effect $A_?$ corresponds to the inconclusive answer ? and $\lambda\in[0;1]$ is such a parameter that $A_?\geq 0$.

\begin{figure}
    \centering
    \includegraphics[scale=0.8]{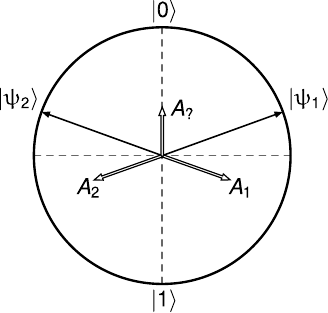}
    \caption{Depiction of an unambiguous state discrimination in the Bloch picture.}
    \label{fig:USDparam}
\end{figure}

In order to analyze this situation, let us parametrize the problem (see also Fig.~\ref{fig:USDparam}). In the qubit case we can always pre-process presented states so that they would be easily described as
\begin{align*}
\ket{\psi_1} &= \cos\omega\ket{0} + \sin\omega \ket{1},\\
\ket{\psi_2} &= \cos\omega\ket{0} - \sin\omega \ket{1},
\end{align*}
with $\omega\in[0;\pi/4]$. The case of $\omega = 0$ corresponds to $\ket{\psi_1}=\ket{\psi_2}$, while $\omega = \pi/4$ describes orthogonal states. We will disregard the pre-processing routine as it is not relevant for this work.

Within this parametrization we have
\begin{align*}
\ket{\psi_1^\perp} &= \sin\omega\ket{0} - \cos\omega \ket{1},\\
\ket{\psi_2^\perp} &= \sin\omega\ket{0} + \cos\omega \ket{1}.
\end{align*}
By minimizing the probability for the inconclusive outcome ? we find  the optimal choice of $\lambda$ to be
\[
\lambda = \frac{1}{2\cos^2\omega}.
\]
We can now explicitly express
\[
A_{1,2} = \frac{1}{2} \begin{pmatrix}
\tan^2\omega & \mp\tan\omega\\
\pm\tan\omega & 1
\end{pmatrix},\quad
A_? = \begin{pmatrix}
1-\tan^2\omega & 0 \\ 0 & 0
\end{pmatrix}.
\]
By construction we see that $A_{1,2}$ are multiples of projectors and we can also observe that $A_?$ is a multiple of a projector, a measurement in the $z$ direction.

An important quantity for us is the probability of inconclusive result,
\[
p_{?} =  p_1 p(A_?|\rho_1) + p_2 p(A_?|\rho_2)
= \frac{1}{2} \tr{A_?P_1} + \frac{1}{2} \tr{A_?P_2} = \cos 2\omega.
\]
The probability of conclusive result (labeled !) is
\[
p_! = 1-p_? = 1-\cos 2\omega = 2\sin^2\omega.
\]

Before analyzing particular sequential measurement scenarios, let us set the notation a bit. In the first case we will consider first the measurement $\B=\{A_?,\id-A_?\}$, where we will denote corresponding outcomes as ? for the inconclusive answer and ! for the conclusive answer. We shall call this measurement \emph{conclusiveness measurement} as it tells us, whether in the subsequent measurement we will obtain a conclusive result or not. In the second case we will start with measurement $\B=\{A_1, \id-A_1\}$, with corresponding outcomes 1 and $1'$ (representing the result `not 1'). We will call this measurement \emph{state 1 measurement} (or for brevity just state measurement) that either tells us whether we were given state $\ket{\psi_1}$ or whether we should continue with the measurement with possibility of obtaining result 2.

\subsection{Conclusiveness measurement as the first measurement}
\label{sec:conclusivemeasurement}

\begin{figure}
    \centering
    a)\includegraphics[scale=0.8]{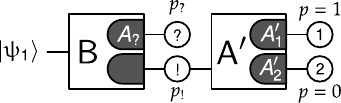}\hskip1cm
    b)\includegraphics[scale=0.8]{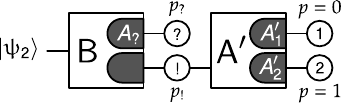}
    \caption{Unambiguous state discrimination where the conclusive measurement is performed first. The two situations show probabilities when a) the state $\ket{\psi_1}$ or b) the state $\ket{\psi_2}$ is on input. Denoted probabilities are conditional, i.e., relating only to the particular measurement.}
    \label{fig:USDconc}
\end{figure}

Let us first look at the symmetric case in which we perform conclusiveness measurement first (see Fig.~\ref{fig:USDconc}) and then perform the outcome measurement. In this case we coarse-grain the measurement $\A=\{A_1,A_2,A_?\}$ by $\B=\{A_?,\id-A_?\}$. The unitaries used in the construction of the coupling from Eq.~(\ref{eq:coupling}) and the basis transformation $U_B$ are determined to be
\[
U_B = \id,\qquad
V_1 = \id,\qquad
V_0 = \begin{pmatrix}
\tan\omega & -\sqrt{1-\tan^2\omega}\\
\sqrt{1-\tan^2\omega} & \tan\omega
\end{pmatrix}.
\]
The pre-measurement state after the coupling transformation is
\begin{equation}
    \label{eq:preMstate}
    \ket{\psi'_{1,2}} = -\sqrt{2}\sin\omega \ket{\pm}\otimes\ket{0} +
    \sqrt{\cos 2\omega}\ket{0}\otimes\ket{1}.
\end{equation}

Denoting $A_!=\id-A_?$, the conclusive result ! is obtained with probability $p(A_!|\psi_1) = p(A_!|\psi_2) = 2\sin^2\omega$ if we measure the ancillary qubit in the state $\ket{0}$. The (normalized) post-measurement states are $\ket{\tilde\psi_1}=\ket{+}$ for the initial state $\ket{\psi_1}$ and $\ket{\tilde\psi_2}=\ket{-}$ for the initial state $\ket{\psi_2}$. These states are orthogonal and, hence, perfectly distinguishable. Indeed, the measurement that shall be performed based on Eq.~(\ref{eq:updatedmeasurement}) is $\A'=\{ P_+, P_- \}$ for the corresponding definitive outcomes 1 and 2; operators $P_\pm$ are projectors into the $\sigma_x$ eigenvectors, i.e., states $\ket{\pm}$ and the measurement can be performed on the system qubit. Altogether we have
\begin{align*}
    p(A_1|\psi_1) &= p(A_!|\psi_1)p(A_1'|\tilde\psi_1)=p(A_!|\psi_1)p(P_+|+)=p(A_!|\psi_1)=2\sin^2\omega,\\
    p(A_1|\psi_2) &= p(A_!|\psi_2)p(A_1'|\tilde\psi_2)=p(A_!|\psi_2)p(P_+|-)=0,\\
    p(A_2|\psi_1) &= p(A_!|\psi_1)p(A_2'|\tilde\psi_1)=p(A_!|\psi_1)p(P_-|+)=0,\\
    p(A_2|\psi_2) &= p(A_!|\psi_2)p(A_2'|\tilde\psi_2)=p(A_!|\psi_2)p(P_-|-)=p(A_!|\psi_2)=2\sin^2\omega.
\end{align*}

As for the inconclusive outcome, the measurement on the ancilla measures the state $\ket{1}$ corresponding to this outcome with probability $p(A_?|\psi_1)=p(A_?|\psi_2)=\cos 2\omega$. The (normalized) post-measurement state is afterward the same for both initial states, $\ket{\tilde\psi_1}=\ket{\tilde\psi_2}=\ket{0}$, and thus loses all the information about the original state (as noted above).

This known result was originally presented in \cite{Peres1988}. The computation serves as a formalized way of obtaining the coupling transformation. Comparing the details of the two results, one finds that $A=2\omega$ and the pre-measurement state from \cite{Peres1988} equals to the state from Eq.~(\ref{eq:preMstate}) up to an unimportant local phase which is due to a slightly different choice of completing unitary $V$ in Eq.~(\ref{eq:halfV}).

\subsection{State measurement as the first measurement}
\label{sec:statemeasurement}

\begin{figure}
    \centering
    a)\includegraphics[scale=0.8]{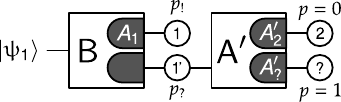}\hskip1cm
    b)\includegraphics[scale=0.8]{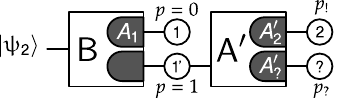}
    \caption{Unambiguous state discrimination where the measurement for the outcome 1 is performed first. The two situations show probabilities when a) the state $\ket{\psi_1}$ or b) the state $\ket{\psi_2}$ is on input. Denoted probabilities are conditional, i.e., relating only to the particular measurement.}
    \label{fig:USDmeas}
\end{figure}

With the framework presented in the previous section, we are able to choose also a different measurement as the first one; it does not have to be the conclusiveness measurement. Let us choose the state $1$ measurement (see Fig.~\ref{fig:USDmeas}), i.e., we want to know whether the presented state $\ket{\psi}$ is $\ket{\psi_1}$. If we are presented the answer 1, we know that the given state was $\ket{\psi_1}$; in the opposite case of the answer labeled as $1'$ (meaning \emph{not 1}) we need to perform a second measurement that shall tell us whether the state was $\ket{\psi_2}$ or the outcome is inconclusive ?.

In the previous case of conclusiveness measurement as the first one to be performed, we saw that in the first step we either got an inconclusive answer, in which case the outcome states for both input states were the same, or we got a conclusive answer, in which case the outcome states for the two input states were orthogonal, and thus perfectly distinguishable.

What can we expect in this scenario, where we first test whether the first state is on the input? Let us assume first, that we got the first state on the input; the test shall then show with probability $p(A_!|\psi_1)$ that we have the state $\ket{\psi_1}$ and with complementary probability $1-p(A_!|\psi_1)=p(A_?|\psi_1)$ we obtain outcome $1'$ and we need to perform the second measurement on the post-measurement state $\ket{\tilde\psi_1}$, where answer that we have the state $\ket{\psi_2}$ has to have zero probability, $p(A_2'|\tilde\psi_1)=0$, and so outcome ? will always be given.

In the case we are presented with the state $\ket{\psi_2}$, the first measurement has to always provide the outcome $1'$, since $p(A_1|\psi_2)=0$; the subsequent measurement on the post-measurement state $\ket{\tilde\psi_2}$ shall afterward lead to the conclusive answer 2 with probability $p(A_2'|\tilde\psi_2)=p(A_2|\psi_2)$ and with probability $p(A_?'|\tilde\psi_2)=p(A_?|\psi_2)$ lead to the inconclusive outcome ?. Let us confirm these expectations.

In this case we coarse-grain the measurement $\A = \{A_1, A_2, A_?\}$ by  $\B = \{A_1, \id - A_1\}$. The unitaries used in the construction of the coupling in Eq.~(\ref{eq:coupling}) and the basis transformation $U_B$ are determined to be
\begin{align*}
    U_B &= \ket{0}\bra{\psi_2^\perp} + \ket{1}\bra{\psi_2} =
    \begin{pmatrix}
    \sin\omega & \cos\omega \\
    \cos\omega & -\sin\omega
    \end{pmatrix},\\
    V_0 &= \frac{1}{\sqrt{2}\cos\omega} \begin{pmatrix}
    \sqrt{\cos 2\omega} & -1 \\
    1 & \sqrt{\cos 2\omega}    
    \end{pmatrix},\\ V_1 &=\id.
\end{align*}

The second measurement $\A'=\{A',\id-A'\}$ is determined based on Eq.~(\ref{eq:updatedmeasurement}). Choosing effect $A'$ to correspond to the outcome 2, and the complementary effect to the outcome ? we obtain
\begin{equation}
\label{eq:updatedmeasurement2}
A'\equiv A_2' = \left(P_2 + \frac{1}{\sqrt{1-\lambda}}P_2^\perp \right)
\lambda P_1^\perp
\left(P_2 + \frac{1}{\sqrt{1-\lambda}}P_2^\perp \right)
\end{equation}

First, supposing we are given the state $\ket{\psi_2}$ on the input, the pre-measurement state is
\[
\ket{\psi'_2} = \ket{1}\otimes\ket{0}.
\]
Since we have $p(A_1|\psi_2)=0$, the ancillary qubit measurement will measure the state $\ket{0}$ with probability $1$; this state corresponds to the outcome $1'$ and we need to follow with the second measurement $\A'$ on the post-measurement state $\ket{\tilde\psi_2}=\ket{\psi_2}$ --- this is the same state as the presented state, as it is the 1-eigenstate of the measurement effect $B=A_1$. The probability that the effect $A'_2$ will be measured in the second measurement, i.e., state 2 will be determined, is directly computed using the Born rule,
\[
p(A_2'|\tilde\psi_2) = \tr{P_2 A'_2} = \lambda\tr{P_1^\perp P_2} = 2\sin^2\omega.
\]
Since we always end up doing this measurement, initial state $\ket{\psi_2}$ leads to the definitive answer 2 with conclusive probability $p(A_2|\psi_2)=2\sin^2\omega$ and to the inconclusive answer ? with probability $p(A_?|\psi_2) = 1 - p(A_2|\psi_2)$.

Now, suppose the state $\ket{\psi_1}$ is presented on the input. Particular computations are more extensive than in the previous case, but still straightforward. The pre-measurement state is
\[
\ket{\psi'_1} = \sqrt{2}\sin\omega\ket{0}\otimes\ket{1}
+ \sqrt{\cos 2\omega}\left(
\sqrt{2}\sin\omega \ket{0} + \sqrt{\cos 2\omega} \ket{1}
\right)\otimes\ket{0}.
\]
In the case state $\ket{1}$ is measured on the ancillary qubit, outcome 1 is assumed and this happens with probability $p(A_1|\psi_1)=2\sin^2\omega$. With probability $p(\id-A_1|\psi_1)=p(A_?|\psi_1)=\cos 2\omega$ outcome $1'$ is provided and we follow with measurement $\A'$. The post-measurement state is
\[
\ket{\tilde\psi_1} = U_B^\dagger\left(
\sqrt{2}\sin\omega \ket{0} + \sqrt{\cos 2\omega} \ket{1}
\right).
\]
After some computation we find that the probability measurement $\A'$ yields outcome 2 is $p(A_2'|\psi_1')=0$ and so the inconclusive outcome ? is always measured.

\subsection{Generalization into biased case}

In the introduction of the problem, we included a possibility to provide states with unequal probabilities. Suppose now that state $\ket{\psi_1}$ is sampled with probability $p$ and the state $\ket{\psi_2}$ with probability $1-p$. Analysis of the situation is similar to the one provided above with the difference that
\[
A_1 = \lambda_1 P_2^\perp,\quad
A_2 = \lambda_2 P_1^\perp,\quad
A_? = \id - A_1 - A_2.
\]
In this case the optimized parameters are $\lambda_1$ and $\lambda_2$, both from the interval $[0;1]$, such that $A_?\geq 0$. In the optimization one minimizes the inconclusive result
\[
p_? = p_1 p(A_?|\psi_1) + p_2 p(A_?|\psi_2)
= p\tr{A_?P_1}+(1-p)\tr{A_?P_2}
\]
or maximizes the success rate $p_{suc} = 1-p_?$.

Denoting the threshold value for $p$ as $p_\theta=\cos^2 2\omega / (1+\cos^2 2\omega)$, the solution is split into two cases, either (i) $p\leq p_\theta$ or $p\geq 1-p_\theta$, or (ii) $p_\theta \leq p \leq 1-p_\theta$.

The case (i) is not relevant for this paper, as in this case one of the states is preferred to such a degree that the most suitable measurement is a projective one for that state, i.e., one of the $\lambda$'s is 0, while the other one is 1. This means that the presented sequential procedure of implementing POVMs is not needed to implement the optimal measurement.

In the other case (ii), all the effects of the measurement are non-trivial, with
\[
    \lambda_1 = \frac{1}{\sin^2 2\omega}\left(1-\cos 2\omega \sqrt{\frac{1-p}{p}}\right),\quad
    \lambda_2 = \frac{1}{\sin^2 2\omega}\left(1-\cos 2\omega \sqrt{\frac{p}{1-p}}\right),
\]
and $A_1$ and $A_2$ are, again, multiples of projectors.
As the optimization procedure for $A_?\geq 0$ is such that one of the eigenvalues $A_?$ is zero, it implies that $A_?$ is a multiple of a projector as well. Hence, all the effects are always multiples of projectors in this task. This means that in both implementations either $V_0=\id$ or $V_1=\id$ and we can disregard corresponding controlled operation.

Furthermore, we can look at the follow-up measurement $\A'$. In both cases, conclusiveness as well as state measurement schemes, this measurement is projective. The reasoning has two steps. In the first step we show that the observed effect $A'$ (which is either $A_1'$ or $A_2'$) is a multiple of a projector. In the second step we show that the multiplication factor has to be 1, i.e., the measurement is projective.

For the first part, the measured effect $A'$ is obtained by a general scheme induced by Eq.~(\ref{eq:updatedmeasurement})
\[
A' = (\id - A_0)^{-1/2} A (\id - A_0)^{-1/2},
\]
where $A'$ is the effect to be observed in the follow-up measurement, $A$ is the intended effect (in our case a multiple of an identity, one of the effects $A_2$ or $A_?$), and $A_0$ is the effect for the definitive outcome from the first measurement (either $A_?$ or $A_1$). In this general notation, we have $A_0=\lambda P_0$ and $A=\mu P$ where $\lambda$ and $\mu$ are some multiplicative constants and $P$ and $P_0$ are projectors. We can see that
\[
    A'\cdot A' = (\id - A_0)^{-1/2} A (\id - A_0)^{-1} A (\id - A_0)^{-1/2} = (\id - A_0)^{-1/2} \kappa A (\id - A_0)^{-1/2} = \kappa A',
\]
where we used that $PQP=\tilde\kappa P$ for projector $P$ and some factor $\tilde\kappa$. We see that $A'\cdot A' = \kappa A'$ and so $A'$ is a multiple of a projector.

In the second step we argue, that $\kappa=1$. If this would not be the case, by denoting $A'=\tau R$ for some $\tau$ and projector $R$ we could exchange the measurement $\A'$ for the projective measurement $\{R, \id - R\}$. This would allow us to obtain higher probabilities for the intended effects (either $A_1$ or $A_2$) which would increase the success rate. Since our measurements are optimal, $\A'$ has to be projective.

\begin{figure}
    \centering
    \input{img/USDconst}
    \caption{Measurement scheme for the (biased) qubit unambiguous state discrimination. Whichever measurement is chosen on the ancilla, the scheme has the same structure, with the first part performing the measurement of the coarse graining. Since the follow-up measurement is always projective, it can be performed on the system qubit by some rotation $W$ into the corresponding basis. Note that this setup does not require classical conditioning inside the quantum circuit, but if the result on the ancilla is for the chosen definitive outcome, the results on the system qubit are disregarded. Also, the ancilla measurement can be postponed so that both measurements are performed at the end.}
    \label{fig:USDconstruction}
\end{figure}

This discussion leads us to a simplified measurement scheme, where the final measurement can be performed on the original system without the need of an ancilla. Altogether, only one ancillary qubit (without intermediate reset) is needed. The situation is depicted in Fig.~\ref{fig:USDconstruction}.

The implementation within the scheme we presented is not expressible in such an easy manner as in the unbiased case, but from the Fig.~\ref{fig:USDconstruction} we see that it can be used in a systematic and computationally accessible way to determine particular elements of the circuit.

\subsection{Usefulness of the POVM implementation scheme}

In the previous section, we have seen that the scheme for the biased unambiguous discrimination of qubit states can be implemented in a very simple way requiring only a single ancillary qubit. The analysis shows that presented scheme can produce circuit settings in a systematic way based on the preferred choice of the measurement on the ancilla. In this case it might not be clear which choice will produce the best result and the factors affecting the decision are numerous and beyond the scope of this paper. However, let us provide a more illustrative usefulness of having possibility of the choice here. We do so be providing yet another generalization of the problem of unambiguous state discrimination beyond the qubit case. Due to the complexity of the problem, we shall provide only estimation of resources without delving into the intricacies of particular computations.

Following the results of \cite{Chefles98, Sedlak09}, unambiguous discrimination is possible only for linearly independent states and, thus, in dimension $d$ we can distinguish unambiguously only up to $d$ states. Together with the inconclusiveness outcome, we need $d+1$ outcomes of the measurement. Translated to qubits, for the implementation of unambiguous state discrimination on $m$ qubits, at least one additional qubit is necessary. This in general is not sufficient, but for our analysis of just a single partial measurement, we need to consider just one ancillary qubit.

In such case one just needs to decide which outcome measurement is to be assigned to the ancillary qubit. A straightforward observation suggests there is space for optimization --- considering Eq.~(\ref{eq:ctrlcoupling}) we see that in the coupling scheme the number of controlled operations equals to the rank of the observed effect. In the standard unambiguous state discrimination \cite{Chefles98} the outcomes (conclusive) effects are multiples of projections and, hence, rank-$1$, while the inconclusive effect $A_?$ has, in general, rank $d-1$. In higher dimension this suggests advantage of assigning one of the conclusive outcomes to the ancilla.

To quantify it, let us restrict to a dimension $d=2^m$ of $m\geq 2$ qubits. The problem of implementing the unambiguous discrimination of $d$ states is split into (i) the part of coupling the system to the ancilla and performing chosen measurement on the ancilla, (ii) performing the rest of the measurement, and finally, (iii) the part for interpreting the result --- if the ancillary measurement provides definitive answer, keep it and discard measurement outcomes on the system, and if the measurement on the ancilla does not provide a definitive answer, deduce correct outcome from the remaining measurements.

We are especially interested in the point (i). Since our aim is to obtain a definitive outcome, choosing the inconclusive outcome for the ancilla seems to be less useful as picking one of the definitive outcomes. For one of the definitive outcomes, which are multiples of a projection, we need to perform only single multi-controlled operation, while for the inconclusive outcome, this can be as many as $d-1$ multi-controlled operations. Two-qubit operations in practical computation bring considerable noise into the computation and the rule of a thumb is to minimize their number. In this case, using results of \cite{vale2023decomposition}, the number of CNOTs to decompose an $m$-qubit multi-controlled real-valued operation would be $16m-24$.

To reduce the number of CNOTs, it is therefore better to perform definitive measurements on the ancilla. This rough estimate minimizes the number of two-qubit operations. However, there can be other factors involving the decision for the ancilla-designated measurement, such as particularities of studied problem, or specific noise profiles of the quantum device.

\section{Discussion}

We have presented a method of transforming complex general quantum measurements into a sequence of simple atomic measurements. Similarly to \cite{AnOi08} we have provided a framework for the analysis, which we used to study quantum unambiguous state discrimination in its simplest setting on a qubit (analytically for the unbiased case and in less detail for the biased case). We were able to show that this framework describes almost exactly the construction of Peres \cite{Peres1988} in the case we perform the conclusiveness measurement first.

However, the framework allows us to choose any other measurement on the ancilla, e.g., the measurement whether we are presented the first state. In such case we devised the measurement procedure. The reason why one might opt for this option might be following. Quantum measurements, even the simple ones, are from the experimental point quite demanding. In the current noisy quantum devices this means that during the measurement we lose a lot of quantum resources (coherence in particular). It might be therefore beneficial to perform measurements that give us the largest amount of information or the most useful information as early in the process of measurement as possible.

Take, for example, a case where we are presented with two states that are close to orthogonal. In such case performing measurement for conclusiveness will be followed by state measurement with high probability, but the state presented to this second measurement will be presented with higher noise due to short coherence times. But if we will perform the measurement for the first state, we will be presented with a definite answer with much higher probability and the second measurement will be less frequently performed. This may lead to a higher overall success rate.

At this point the procedure might not seem very useful, as in this simplest setting the measurement on the original system can be performed irrespective of the measurement on the qubit system. However, in constructions of measurements with larger number of outcomes, this flexibility might become important, as the measurements to be performed will be conditioned on the previous outcomes. The situation becomes apparent already for $n=4$ outcomes, where subsequent measurements depend on the previous outcome.

This paper, however, does not present results on which choice for the ancillary measurement is optimal; it offers only a systematic construction based on the choice. The depth of the circuit may depend, in addition to the choice of the ancillary measurement, also on many other factors, such as the form of subsequent measurement, the ability to efficiently decompose different parts of implementation circuit, or even technical parameters of the particular quantum device.

We hope that this procedure can offer us more precise measurement processes on current quantum devices by offering a flexible and generic approach to implementing POVMs.

\section*{Acknowledgement}

The authors acknowledge support from GRUPIK (MUNI/G/1211/2017) project, and VEGA project \emph{HOQIT} No.~2/0161/19. This research was made possible by funding from
QuantERA, an ERA-Net cofund in Quantum Technologies, under the project eDICT. D.R.~thanks Michal Sedl\'ak for discussions.

\section*{Data availability}

All data generated or analysed during this study are included in this published article.



\begin{thebibliography}{10}
\expandafter\ifx\csname url\endcsname\relax
  \def\url#1{\burl{#1}}\fi
\expandafter\ifx\csname urlprefix\endcsname\relax\def\urlprefix{URL }\fi
\providecommand{\bibinfo}[2]{#2}
\providecommand{\eprint}[2][]{\url{#2}}
\providecommand{\doi}[1]{\url{https://doi.org/#1}}
\bibcommenthead

\bibitem{Peres2006}
\bibinfo{author}{Peres, A.}
\newblock \emph{\bibinfo{title}{Quantum Theory: Concepts and Methods}}
  Fundamental Theories of Physics (\bibinfo{publisher}{Springer Netherlands},
  \bibinfo{year}{2006}).
\newblock \urlprefix\url{https://books.google.cz/books?id=pQXSBwAAQBAJ}.

\bibitem{Renes2003}
\bibinfo{author}{Renes, J.~M.}, \bibinfo{author}{Blume-Kohout, R.},
  \bibinfo{author}{Scott, A.~J.} \& \bibinfo{author}{Caves, C.~M.}
\newblock \bibinfo{title}{Symmetric informationally complete quantum
  measurements}.
\newblock \emph{\bibinfo{journal}{Journal of Mathematical Physics}}
  \textbf{\bibinfo{volume}{45}}, \bibinfo{pages}{2171--2180}
  (\bibinfo{year}{2004}).

\bibitem{Bergou2010}
\bibinfo{author}{Bergou, J.~A.}
\newblock \bibinfo{title}{Discrimination of quantum states}.
\newblock \emph{\bibinfo{journal}{Journal of Modern Optics}}
  \textbf{\bibinfo{volume}{57}}, \bibinfo{pages}{160--180}
  (\bibinfo{year}{2010}).

\bibitem{Derka1998}
\bibinfo{author}{Derka, R.}, \bibinfo{author}{Bu\v{z}ek, V.} \&
  \bibinfo{author}{Ekert, A.~K.}
\newblock \bibinfo{title}{Universal algorithm for optimal estimation of quantum
  states from finite ensembles via realizable generalized measurement}.
\newblock \emph{\bibinfo{journal}{Phys. Rev. Lett.}}
  \textbf{\bibinfo{volume}{80}}, \bibinfo{pages}{1571--1575}
  (\bibinfo{year}{1998}).
\newblock \urlprefix\url{https://link.aps.org/doi/10.1103/PhysRevLett.80.1571}.

\bibitem{Bennet2002}
\bibinfo{author}{Bennett, C.~H.}
\newblock \bibinfo{title}{Quantum cryptography using any two nonorthogonal
  states}.
\newblock \emph{\bibinfo{journal}{Phys. Rev. Lett.}}
  \textbf{\bibinfo{volume}{68}}, \bibinfo{pages}{3121--3124}
  (\bibinfo{year}{1992}).
\newblock \urlprefix\url{https://link.aps.org/doi/10.1103/PhysRevLett.68.3121}.

\bibitem{Renes04Crypt}
\bibinfo{author}{Renes, J.~M.}
\newblock \bibinfo{title}{Spherical-code key-distribution protocols for
  qubits}.
\newblock \emph{\bibinfo{journal}{Phys. Rev. A}} \textbf{\bibinfo{volume}{70}},
  \bibinfo{pages}{052314} (\bibinfo{year}{2004}).
\newblock \urlprefix\url{https://link.aps.org/doi/10.1103/PhysRevA.70.052314}.

\bibitem{Nielsen2010}
\bibinfo{author}{Nielsen, M.} \& \bibinfo{author}{Chuang, I.}
\newblock \emph{\bibinfo{title}{Quantum Computation and Quantum Information:
  10th Anniversary Edition}}  (\bibinfo{publisher}{Cambridge University Press},
  \bibinfo{year}{2010}).

\bibitem{Jozsa2003}
\bibinfo{author}{Jozsa, R.} \emph{et~al.}
\newblock \bibinfo{title}{Entanglement cost of generalised measurements}.
\newblock \emph{\bibinfo{journal}{Quantum Inf. Comput.}}
  \textbf{\bibinfo{volume}{3}}, \bibinfo{pages}{405--422}
  (\bibinfo{year}{2003}).
\newblock \urlprefix\url{http://portal.acm.org/citation.cfm?id=2011546}.

\bibitem{Gisin96}
\bibinfo{author}{Gisin, N.}
\newblock \bibinfo{title}{Hidden quantum nonlocality revealed by local
  filters}.
\newblock \emph{\bibinfo{journal}{Physics Letters A}}
  \textbf{\bibinfo{volume}{210}}, \bibinfo{pages}{151 -- 156}
  (\bibinfo{year}{1996}).
\newblock
  \urlprefix\url{http://www.sciencedirect.com/science/article/pii/S0375960196800016}.

\bibitem{Vertesi2010}
\bibinfo{author}{V\'ertesi, T.} \& \bibinfo{author}{Bene, E.}
\newblock \bibinfo{title}{Two-qubit bell inequality for which positive
  operator-valued measurements are relevant}.
\newblock \emph{\bibinfo{journal}{Phys. Rev. A}} \textbf{\bibinfo{volume}{82}},
  \bibinfo{pages}{062115} (\bibinfo{year}{2010}).
\newblock \urlprefix\url{https://link.aps.org/doi/10.1103/PhysRevA.82.062115}.

\bibitem{tavakoli2021}
\bibinfo{author}{Tavakoli, A.}, \bibinfo{author}{Farkas, M.},
  \bibinfo{author}{Rosset, D.}, \bibinfo{author}{Bancal, J.-D.} \&
  \bibinfo{author}{Kaniewski, J.}
\newblock \bibinfo{title}{Mutually unbiased bases and symmetric informationally
  complete measurements in bell experiments}.
\newblock \emph{\bibinfo{journal}{Science Advances}}
  \textbf{\bibinfo{volume}{7}}, \bibinfo{pages}{eabc3847}
  (\bibinfo{year}{2021}).
\newblock
  \urlprefix\url{https://www.science.org/doi/abs/10.1126/sciadv.abc3847}.

\bibitem{Acin2016}
\bibinfo{author}{Ac\'in, A.}, \bibinfo{author}{Pironio, S.},
  \bibinfo{author}{V\'ertesi, T.} \& \bibinfo{author}{Wittek, P.}
\newblock \bibinfo{title}{Optimal randomness certification from one entangled
  bit}.
\newblock \emph{\bibinfo{journal}{Phys. Rev. A}} \textbf{\bibinfo{volume}{93}},
  \bibinfo{pages}{040102} (\bibinfo{year}{2016}).
\newblock \urlprefix\url{https://link.aps.org/doi/10.1103/PhysRevA.93.040102}.

\bibitem{Gomez2016}
\bibinfo{author}{G\'omez, E.~S.} \emph{et~al.}
\newblock \bibinfo{title}{Device-independent certification of a nonprojective
  qubit measurement}.
\newblock \emph{\bibinfo{journal}{Phys. Rev. Lett.}}
  \textbf{\bibinfo{volume}{117}}, \bibinfo{pages}{260401}
  (\bibinfo{year}{2016}).
\newblock
  \urlprefix\url{https://link.aps.org/doi/10.1103/PhysRevLett.117.260401}.

\bibitem{Preskill18}
\bibinfo{author}{Preskill, J.}
\newblock \bibinfo{title}{Quantum computing in the nisq era and beyond}.
\newblock \emph{\bibinfo{journal}{Quantum}} \textbf{\bibinfo{volume}{2}},
  \bibinfo{pages}{79} (\bibinfo{year}{2018}).
\newblock \urlprefix\url{https://doi.org/10.22331/q-2018-08-06-79}.

\bibitem{Andersson12b}
\bibinfo{author}{Andersson, E.}
\newblock \bibinfo{title}{Optimal minimum-cost quantum measurements for
  imperfect detection}.
\newblock \emph{\bibinfo{journal}{Phys. Rev. A}} \textbf{\bibinfo{volume}{86}},
  \bibinfo{pages}{012120} (\bibinfo{year}{2012}).
\newblock \urlprefix\url{https://link.aps.org/doi/10.1103/PhysRevA.86.012120}.

\bibitem{MaZiOs20}
\bibinfo{author}{Maciejewski, F.~B.}, \bibinfo{author}{Zimbor{\'{a}}s, Z.} \&
  \bibinfo{author}{Oszmaniec, M.}
\newblock \bibinfo{title}{Mitigation of readout noise in near-term quantum
  devices by classical post-processing based on detector tomography}.
\newblock \emph{\bibinfo{journal}{{Quantum}}} \textbf{\bibinfo{volume}{4}},
  \bibinfo{pages}{257} (\bibinfo{year}{2020}).
\newblock \urlprefix\url{https://doi.org/10.22331/q-2020-04-24-257}.

\bibitem{OsGuWiAc17}
\bibinfo{author}{Oszmaniec, M.}, \bibinfo{author}{Guerini, L.},
  \bibinfo{author}{Wittek, P.} \& \bibinfo{author}{Ac\'{\i}n, A.}
\newblock \bibinfo{title}{Simulating positive-operator-valued measures with
  projective measurements}.
\newblock \emph{\bibinfo{journal}{Phys. Rev. Lett.}}
  \textbf{\bibinfo{volume}{119}}, \bibinfo{pages}{190501}
  (\bibinfo{year}{2017}).
\newblock
  \urlprefix\url{https://link.aps.org/doi/10.1103/PhysRevLett.119.190501}.

\bibitem{OsMaPu19}
\bibinfo{author}{Oszmaniec, M.}, \bibinfo{author}{Maciejewski, F.~B.} \&
  \bibinfo{author}{Pucha\l{}a, Z.}
\newblock \bibinfo{title}{Simulating all quantum measurements using only
  projective measurements and postselection}.
\newblock \emph{\bibinfo{journal}{Phys. Rev. A}}
  \textbf{\bibinfo{volume}{100}}, \bibinfo{pages}{012351}
  (\bibinfo{year}{2019}).
\newblock \urlprefix\url{https://link.aps.org/doi/10.1103/PhysRevA.100.012351}.

\bibitem{AnOi08}
\bibinfo{author}{Andersson, E.} \& \bibinfo{author}{Oi, D. K.~L.}
\newblock \bibinfo{title}{Binary search trees for generalized measurements}.
\newblock \emph{\bibinfo{journal}{Phys. Rev. A}} \textbf{\bibinfo{volume}{77}},
  \bibinfo{pages}{052104} (\bibinfo{year}{2008}).
\newblock \urlprefix\url{https://link.aps.org/doi/10.1103/PhysRevA.77.052104}.

\bibitem{SiMaOs2021}
\bibinfo{author}{Singal, T.}, \bibinfo{author}{Maciejewski, F.~B.} \&
  \bibinfo{author}{Oszmaniec, M.}
\newblock \bibinfo{title}{Implementation of quantum measurements using
  classical resources and only a single ancillary qubit}.
\newblock \emph{\bibinfo{journal}{npj Quantum Information}}
  \textbf{\bibinfo{volume}{8}}, \bibinfo{pages}{82} (\bibinfo{year}{2022}).

\bibitem{AnOiKaBrAn10}
\bibinfo{author}{Anders, J.}, \bibinfo{author}{Oi, D. K.~L.},
  \bibinfo{author}{Kashefi, E.}, \bibinfo{author}{Browne, D.~E.} \&
  \bibinfo{author}{Andersson, E.}
\newblock \bibinfo{title}{Ancilla-driven universal quantum computation}.
\newblock \emph{\bibinfo{journal}{Phys. Rev. A}} \textbf{\bibinfo{volume}{82}},
  \bibinfo{pages}{020301} (\bibinfo{year}{2010}).
\newblock \urlprefix\url{https://link.aps.org/doi/10.1103/PhysRevA.82.020301}.

\bibitem{ShenEtal17}
\bibinfo{author}{Shen, C.} \emph{et~al.}
\newblock \bibinfo{title}{Quantum channel construction with circuit quantum
  electrodynamics}.
\newblock \emph{\bibinfo{journal}{Phys. Rev. B}} \textbf{\bibinfo{volume}{95}},
  \bibinfo{pages}{134501} (\bibinfo{year}{2017}).
\newblock \urlprefix\url{https://link.aps.org/doi/10.1103/PhysRevB.95.134501}.

\bibitem{HeZi11}
\bibinfo{author}{Heinosaari, T.} \& \bibinfo{author}{Ziman, M.}
\newblock \emph{\bibinfo{title}{The Mathematical Language of Quantum Theory:
  From Uncertainty to Entanglement}}  (\bibinfo{publisher}{Cambridge University
  Press}, \bibinfo{year}{2011}).

\bibitem{RoDeMaGi17}
\bibinfo{author}{Rosati, M.}, \bibinfo{author}{De~Palma, G.},
  \bibinfo{author}{Mari, A.} \& \bibinfo{author}{Giovannetti, V.}
\newblock \bibinfo{title}{Optimal quantum state discrimination via nested
  binary measurements}.
\newblock \emph{\bibinfo{journal}{Phys. Rev. A}} \textbf{\bibinfo{volume}{95}},
  \bibinfo{pages}{042307} (\bibinfo{year}{2017}).
\newblock \urlprefix\url{https://link.aps.org/doi/10.1103/PhysRevA.95.042307}.

\bibitem{GaReSa23}
\bibinfo{author}{Galvis-Florez, C.~A.}, \bibinfo{author}{Reitzner, D.} \&
  \bibinfo{author}{Särkkä, S.}
\newblock \bibinfo{editor}{\;} (ed.) \emph{\bibinfo{title}{Single qubit state
  estimation on nisq devices with limited resources and sic-povms}}.
\newblock (ed.\bibinfo{editor}{\;}) \emph{\bibinfo{booktitle}{2023 IEEE
  International Conference on Quantum Computing and Engineering (QCE)}},
  Vol.~\bibinfo{volume}{01}, \bibinfo{pages}{111--119} (\bibinfo{year}{2023}).

\bibitem{YoBa19}
\bibinfo{author}{Yordanov, Y.~S.} \& \bibinfo{author}{Barnes, C. H.~W.}
\newblock \bibinfo{title}{Implementation of a general single-qubit positive
  operator-valued measure on a circuit-based quantum computer}.
\newblock \emph{\bibinfo{journal}{Phys. Rev. A}}
  \textbf{\bibinfo{volume}{100}}, \bibinfo{pages}{062317}
  (\bibinfo{year}{2019}).
\newblock \urlprefix\url{https://link.aps.org/doi/10.1103/PhysRevA.100.062317}.

\bibitem{Fischer17}
\bibinfo{author}{Fischer, L.~E.} \emph{et~al.}
\newblock \bibinfo{title}{Ancilla-free implementation of generalized
  measurements for qubits embedded in a qudit space}.
\newblock \emph{\bibinfo{journal}{Phys. Rev. Res.}}
  \textbf{\bibinfo{volume}{4}}, \bibinfo{pages}{033027} (\bibinfo{year}{2022}).
\newblock
  \urlprefix\url{https://link.aps.org/doi/10.1103/PhysRevResearch.4.033027}.

\bibitem{PiZaBaMa23}
\bibinfo{author}{Pinto, D.~F.}, \bibinfo{author}{Zanetti, M.~S.},
  \bibinfo{author}{Basso, M. L.~W.} \& \bibinfo{author}{Maziero, J.}
\newblock \bibinfo{title}{Simulation of positive operator-valued measures and
  quantum instruments via quantum state-preparation algorithms}.
\newblock \emph{\bibinfo{journal}{Phys. Rev. A}}
  \textbf{\bibinfo{volume}{107}}, \bibinfo{pages}{022411}
  (\bibinfo{year}{2023}).
\newblock \urlprefix\url{https://link.aps.org/doi/10.1103/PhysRevA.107.022411}.

\bibitem{AhPa05}
\bibinfo{author}{Ahnert, S.~E.} \& \bibinfo{author}{Payne, M.~C.}
\newblock \bibinfo{title}{General implementation of all possible
  positive-operator-value measurements of single-photon polarization states}.
\newblock \emph{\bibinfo{journal}{Phys. Rev. A}} \textbf{\bibinfo{volume}{71}},
  \bibinfo{pages}{012330} (\bibinfo{year}{2005}).
\newblock \urlprefix\url{https://link.aps.org/doi/10.1103/PhysRevA.71.012330}.

\bibitem{AhPa06}
\bibinfo{author}{Ahnert, S.~E.} \& \bibinfo{author}{Payne, M.~C.}
\newblock \bibinfo{title}{All possible bipartite positive-operator-value
  measurements of two-photon polarization states}.
\newblock \emph{\bibinfo{journal}{Phys. Rev. A}} \textbf{\bibinfo{volume}{73}},
  \bibinfo{pages}{022333} (\bibinfo{year}{2006}).
\newblock \urlprefix\url{https://link.aps.org/doi/10.1103/PhysRevA.73.022333}.

\bibitem{OtAsNo12}
\bibinfo{author}{Ota, Y.}, \bibinfo{author}{Ashhab, S.} \&
  \bibinfo{author}{Nori, F.}
\newblock \bibinfo{title}{Implementing general measurements on linear optical
  and solid-state qubits}.
\newblock \emph{\bibinfo{journal}{Phys. Rev. A}} \textbf{\bibinfo{volume}{85}},
  \bibinfo{pages}{043808} (\bibinfo{year}{2012}).
\newblock \urlprefix\url{https://link.aps.org/doi/10.1103/PhysRevA.85.043808}.

\bibitem{DrBrKo14}
\bibinfo{author}{Dressel, J.}, \bibinfo{author}{Brun, T.~A.} \&
  \bibinfo{author}{Korotkov, A.~N.}
\newblock \bibinfo{title}{Implementing generalized measurements with
  superconducting qubits}.
\newblock \emph{\bibinfo{journal}{Phys. Rev. A}} \textbf{\bibinfo{volume}{90}},
  \bibinfo{pages}{032302} (\bibinfo{year}{2014}).
\newblock \urlprefix\url{https://link.aps.org/doi/10.1103/PhysRevA.90.032302}.

\bibitem{Bian15}
\bibinfo{author}{Bian, Z.} \emph{et~al.}
\newblock \bibinfo{title}{Realization of single-qubit positive-operator-valued
  measurement via a one-dimensional photonic quantum walk}.
\newblock \emph{\bibinfo{journal}{Phys. Rev. Lett.}}
  \textbf{\bibinfo{volume}{114}}, \bibinfo{pages}{203602}
  (\bibinfo{year}{2015}).
\newblock
  \urlprefix\url{https://link.aps.org/doi/10.1103/PhysRevLett.114.203602}.

\bibitem{Zhao15}
\bibinfo{author}{Zhao, Y.-y.} \emph{et~al.}
\newblock \bibinfo{title}{Experimental realization of generalized qubit
  measurements based on quantum walks}.
\newblock \emph{\bibinfo{journal}{Phys. Rev. A}} \textbf{\bibinfo{volume}{91}},
  \bibinfo{pages}{042101} (\bibinfo{year}{2015}).
\newblock \urlprefix\url{https://link.aps.org/doi/10.1103/PhysRevA.91.042101}.

\bibitem{SiArSa22}
\bibinfo{author}{Singh, J.}, \bibinfo{author}{Arvind} \&
  \bibinfo{author}{Goyal, S.~K.}
\newblock \bibinfo{title}{Implementation of discrete positive operator valued
  measures on linear optical systems using cosine-sine decomposition}.
\newblock \emph{\bibinfo{journal}{Phys. Rev. Res.}}
  \textbf{\bibinfo{volume}{4}}, \bibinfo{pages}{013007} (\bibinfo{year}{2022}).
\newblock
  \urlprefix\url{https://link.aps.org/doi/10.1103/PhysRevResearch.4.013007}.

\bibitem{Dieks1988}
\bibinfo{author}{Dieks, D.}
\newblock \bibinfo{title}{Overlap and distinguishability of quantum states}.
\newblock \emph{\bibinfo{journal}{Physics Letters A}}
  \textbf{\bibinfo{volume}{126}}, \bibinfo{pages}{303 -- 306}
  (\bibinfo{year}{1988}).
\newblock
  \urlprefix\url{http://www.sciencedirect.com/science/article/pii/0375960188908407}.

\bibitem{Ivanovic1987}
\bibinfo{author}{Ivanovic, I.}
\newblock \bibinfo{title}{How to differentiate between non-orthogonal states}.
\newblock \emph{\bibinfo{journal}{Physics Letters A}}
  \textbf{\bibinfo{volume}{123}}, \bibinfo{pages}{257 -- 259}
  (\bibinfo{year}{1987}).
\newblock
  \urlprefix\url{http://www.sciencedirect.com/science/article/pii/0375960187902222}.

\bibitem{Peres1988}
\bibinfo{author}{Peres, A.}
\newblock \bibinfo{title}{How to differentiate between non-orthogonal states}.
\newblock \emph{\bibinfo{journal}{Physics Letters A}}
  \textbf{\bibinfo{volume}{128}}, \bibinfo{pages}{19} (\bibinfo{year}{1988}).
\newblock
  \urlprefix\url{http://www.sciencedirect.com/science/article/pii/0375960188910341}.

\bibitem{Chefles98}
\bibinfo{author}{Chefles, A.}
\newblock \bibinfo{title}{Unambiguous discrimination between linearly
  independent quantum states}.
\newblock \emph{\bibinfo{journal}{Physics Letters A}}
  \textbf{\bibinfo{volume}{239}}, \bibinfo{pages}{339} (\bibinfo{year}{1998}).

\bibitem{Sedlak09}
\bibinfo{author}{Sedl\'ak, M.}
\newblock \bibinfo{title}{Quantum theory of unambiguous measurements}.
\newblock \emph{\bibinfo{journal}{ACTA PHYSICA SLOVACA}}
  \textbf{\bibinfo{volume}{59}}, \bibinfo{pages}{653} (\bibinfo{year}{2009}).
\newblock
  \urlprefix\url{http://www.citebase.org/abstract?id=oai:arXiv.org:1003.2448}.

\bibitem{vale2023decomposition}
\bibinfo{author}{Vale, R.}, \bibinfo{author}{Azevedo, T. M.~D.},
  \bibinfo{author}{Araújo, I. C.~S.}, \bibinfo{author}{Araujo, I.~F.} \&
  \bibinfo{author}{da~Silva, A.~J.}
\newblock \bibinfo{title}{Decomposition of multi-controlled special unitary
  single-qubit gates} (\bibinfo{year}{2023}).
\newblock \eprint{2302.06377}.

\end{thebibliography}
\end{document}